\documentclass[useAMS,usenatbib]{mn2e}
\usepackage{graphicx}
\title[Gas dynamics in two quasars at z$\ge$3.5]{The dynamics of the
ionized and molecular ISM in powerful obscured quasars at
z$\ge$3.5\thanks{Based on observations carried out with the Very Large
Telescope of ESO under Program ID 084.B-0161 and with the Plateau de
Bure Interferometer of IRAM under Program ID T040.}}
\author[Nesvadba et al.]{\parbox[h]{\textwidth}{N.P.H
Nesvadba$^{1}$\thanks{E-mail: nicole.nesvadba@ias.u-psud.fr},
M. Polletta$^{2}$, M.~D.~Lehnert$^{3}$, J.~Bergeron$^{4}$,
C.~De~Breuck$^{5}$, G.~Lagache $^{1}$ and A.~Omont$^{4}$}\\
$^{1}$Institut d'Astrophysique Spatiale, CNRS,
    Universit\'e Paris-Sud, Bat. 120-121, 91405 Orsay, France\\
$^{2}$INAF - IASF Milano, via E. Bassini, 20133, Italy\\
$^{3}$GEPI, Observatoire de Paris-Meudon, 5 place Jules Janssen, 92195 Meudon, France\\
$^{4}$Institut d'Astrophysique de Paris, CNRS \& Universite Pierre et
Marie Curie, 98bis, bd Arago, 75014 Paris, France\\
$^{5}$European Southern Observatory, Karl-Schwarzschild Strasse, Garching bei M\"unchen,
Germany}

\begin{document}

\date{Accepted . Received ; in original form }


\maketitle

\label{firstpage}

\begin{abstract}
We present an analysis of the kinematics and excitation of the warm
ionized gas in two obscured, powerful quasars at z$\ge$3.5 from the
SWIRE survey, SWIRE~J022513.90-043419.9 and SWIRE~J022550.67-042142,
based on imaging spectroscopy on the VLT. Line ratios in both targets
are consistent with luminous narrow-line regions of AGN.
SWIRE~J022550.67-042142 has very broad (FWHM$=$2000 km s$^{-1}$),
spatially compact [OIII] line emission. SWIRE~J022513.90-043419.9 is
spatially resolved, has complex line profiles of H$\beta$ and [OIII],
including broad wings with blueshifts of up to $-$1500 km s$^{-1}$
relative to the narrow [OIII]$\lambda$5007 component, and widths of up
to FWHM$=$5000 km s$^{-1}$. Estimating the systemic redshift from the
narrow H$\beta$ line, as is standard for AGN host galaxies,
implies that a significant fraction of the molecular gas 
is blueshifted relative to the
systemic velocity. Thus the molecular gas could be participating in
the outflow. Significant fractions of the ionized and molecular gas
reach velocities greater than the escape velocity. We compare empirical
and modeling constraints for different energy injection mechanisms,
such as merging, star formation, and momentum-driven AGN winds. We
argue that the radio source is the most likely culprit, in spite of
the sources rather modest radio power of $10^{25}$ W Hz$^{-1}$. Such a radio
power is not uncommon for intense starburst galaxies at z$\sim$2. We
disucss these results in light of the co-evolution of AGN and their
host galaxy.
\end{abstract}


\section{Introduction}
\label{sec:introduction}

Massive galaxies provide important constraints for our understanding
of galaxy evolution.  Observational and theoretical progress has led
to a reasonably clear picture: Massive galaxies formed most of their
stars in relatively short, intense bursts at high redshift, that were
triggered by major, gas-rich mergers, and regulated by the energy
injection from star formation and active galactic nuclei
\citep[e.g.,][]{hopkins06, granato06, chakrabarti08,
narayanan08}. This basic scenario is simple and elegant, and broadly
consistent with ensemble studies of statistical samples of galaxies
with intense star formation and powerful AGN. However, given the
complexity of these processes, neither ensemble studies nor models
alone can show conclusively if the different components of this model
-- merger, star formation and AGN activity -- interact as postulated,
and if this has the predicted impact on the interstellar medium.  This
is particularly true for the role of the energy injection by the AGN,
for which we do not yet have a good physical understanding.

The most direct way of overcoming these limitations is through
detailed observations of the gas kinematics and energetics in massive
galaxies that are in key phases along this sequence. Ideally, such an
approach must focus on an {\it in situ} study of massive galaxies
during their major phase of growth -- hence we must observe galaxies
at high redshift.

Here we present an analysis of the warm ionized interstellar medium in
two such galaxies, which is based on deep rest-frame optical
integral-field spectroscopy obtained with the Very Large Telescope of
ESO. SWIRE~J022513.90-043419.9 and SWIRE~J022550.67-042142.2 (SW022513
and SW022550 hereafter) are two obscured quasars at z$\ge$3.5
\citep{polletta08} and are the two most luminous 24$\mu$m emitters at
high redshift in the SWIRE survey \citep{lonsdale03}. We also include
the results of a recent CO(4--3) analysis into our discussion, which
has been presented by \citet{polletta10}.

\citet{polletta08} presented a detailed analysis of the
multi-wavelength photometric properties of these galaxies. Both
galaxies host powerful starbursts (${\cal L}\sim 10^{13} {\cal
L}_{\odot}$) and luminous, obscured quasars with bolometric
luminosities $> 10^{13} {\cal L}_{\odot}$. Centimeter radio
observations suggest the presence of moderately powerful radio sources
of order $10^{25}$ W Hz$^{-1}$ at 1.4 GHz in both targets, which
appear too powerful to be entirely powered by star formation. The
rest-UV spectrum of SW022550 shows bright, high-ionization emission
lines, in particular NV$\lambda$1240, and weak UV continuum emission,
which are the typical signatures of luminous, obscured quasars
\citep{polletta08}. SW022513 is well detected in the X-ray
\citep{pierre07} with a luminosity, ${\cal L}$(2-10 keV) = $8.2\times
10^{44}$ erg s$^{-1}$ in the hard X-ray that is 7 times brigher than
the soft X-ray luminosity at 0.5-2 keV.  It is thus a luminous
absorbed X-ray source, e.g. a factor 3 brighter than the archetypal
type-2 quasar CDFS-202 at z$=$3.700 \citep{norman02}. SW022513 is most
probably Compton-thick since its hard X-ray band luminosity is about
only 1/50 of that measured around 24$\mu$m.

\citet{polletta08} also present low-resolution ISAAC longslit
spectroscopy in the near-infrared for both targets, and rest-frame UV
spectroscopy for SW022550, showing that both sources are luminous line
emitters.  Recent IRAM Plateau de Bure millimeter interferometry of
the CO(4--3) line shows that both galaxies are luminous CO line
emitters with ${\cal L}_{CO}\sim 5\times 10^{10} {\cal L}_{\odot}$ K
km s$^{-1}$ pc$^2$ each, corresponding to $4\times 10^{10}$
M$_{\odot}$ of cold molecular gas \citep{polletta10}.  SW022550 has a
broad, double-horned CO line profile, whereas the CO(4--3) line in
SW022513 is also broad, FWHM$=$1000 km s$^{-1}$, and featureless, and
is not very well fit with a Gaussian profile. Short gas consumption
timescales suggest the galaxies may be near the end of their epoch of
intense star formation, which is when galaxy evolution models
postulate the impact of the AGN should be greatest \citep{springel05,
narayanan08}.

Targeting particularly powerful quasars is important to isolate the
impact of the AGN from that of the starburst (or potentially a
merger). By contrasting our targets and galaxies that have equally
intense star formation, but less powerful AGN we may hope
to identify the unique signatures of the AGN. In addition we may set
upper limits on what impact a luminous, obscured AGN may possibly have
on the gas kinematics of its host galaxy. Both aspects are important
to quantify the impact of AGN on the evolution of their host galaxies,
as they may serve as benchmarks to develop analyses of less powerful,
more frequent forms of feedback later on -- and to infer whether these
are indeed observationally feasible with present-day instruments.

Both galaxies have kpc-sized narrow-line regions (NLRs) that have
strongly disturbed gas kinematics and forbidden line emission with
widths of up to few 1000 km s$^{-1}$. SW022513 has complex line
profiles with broad, blueshifted components seen in
[OIII]$\lambda\lambda$4959,5007 and H$\beta$. These are qualitatively
similar, but broader than the blue wing previously found by
\citet{alexander10} in the submillimeter-selected quasar
SMMJ1237+6203. Our analysis focuses on the gas kinematics, including
the ionized and molecular gas studied by \citet{polletta10}. We
compare with the energy output of SW022513 and SW022550 through AGN
and star formation and discuss what physical processes may contribute
to driving such an outflow.

Throughout our analysis we adopt a flat cosmology where H$_0$ = 70 km
s$^{-1}$ Mpc$^{-1}$, $\Omega_{\Lambda}$=0.7, $\Omega_{M}=0.3$. With
this cosmology, the luminosity distance to SW022513 at z=3.42 is 29.7
Gpc. 1 arcsec corresponds to a projected distance of 7.4 kpc. SW022550
is at a luminosity distance D$_L=$34.4 Gpc, and 1 arcsec corresponds
to a projected distance of 7.0 kpc. 

\section{Observations and data reduction}
\label{sec:observations} We observed both targets with the SINFONI
integral-field spectrograph on the VLT. SINFONI is an image slicer
which gives a field of view of 8\arcsec$\times$8\arcsec\ with a pixel
scale of~250 mas in the seeing-limited mode. We used the H$+$K grating
which covers the near-infrared H and K bands simultaneously from
1.45$\mu$m to 2.45$\mu$m at a spectral resolving power of R$\sim$1500
($\sim$200 km s$^{-1}$).

Both targets were observed in October and November 2009 under good and
stable atmospheric conditions. The program was carried out in service
mode under Program ID 384.B-0161. We obtained 3 hrs of on-source
observing time per target, split into individual exposures of 300
seconds to allow for a good sky subtraction. Since we expected that
both sources were smaller than the field of view we adopted a dither
pattern where the sources fall onto different parts of the detector,
but within the field of view at all times during
observations. This allows to use one exposure to subtract the sky from
the subsequent exposure, and makes taking dedicated sky frames
unnessesary.

Data reduction relies on the standard IRAF tools to reduce longslit
spectra \citep{tody93}, which we modified to meet the special
requirements of integral-field spectroscopy. Our data reduction has
been extensively described elsewhere \citep[e.g.,][]{nesvadba06b,
nesvadba07b, nesvadba08b}. The absolute flux scale was determined from
standard star observations at the end of each one-hour long observing
sequence, and we used the same stars to measure the seeing, which
typically has a full-width-at-half-maximum of 0.5\arcsec\ to 0.6\arcsec. 

\section{Results}
\label{sec:results}

\subsection{Continuum emission}
\label{ssec:continuum}
We detect continuum emission at the position expected from the IRAC
3.6$\mu$m imaging in both galaxies. The spectral coverage of our H$+$K
data corresponds to 3300\AA\ to 5500\AA\ and to 2900\AA\ to 5000\AA\
in the rest-frame of SW022513 and SW022550, respectively. The
continuum is relatively faint in both galaxies, but nonetheless
detected at a significance of 1$-$2$\sigma$ per pixel in each spectral
bin of SW022513 and SW022550. To obtain a more robust measurement, we
collapsed both cubes at wavelengths without strong line emission, and
detected the continuum at significances of $>$3$\sigma$ and up to
14$\sigma$ per spatial pixel in the collapsed image over the area of
one PSF.  We did not attempt to constrain the overall spectral shape
of the continuum, however notice that it seems to have a blue slope in
both targets broadly consistent with originating from the AGN.  In
this case, the continuum could be either produced by scattered or
direct light from the AGN, or represent nebular continuum
\citep[e.g.,][]{vernet01}, A more detailed discussion of the continuum
emission is beyond the scope of this paper.

The continuum of SW022550 is compact, in SW022513 it is marginally
extended along the North-South axis at $\sim 3\sigma$ significance (per
spatial pixel, see Figure~\ref{fig:SW022513_maps}). We also detect
very faint continuum emission associated with the faint line emission
north from the nucleus in SW022513. 

\subsection{Emission-line gas}
\label{ssec:emissionlinegas}
\subsubsection{SW022513}

\begin{figure} \centering
\includegraphics[width=0.7\textwidth]{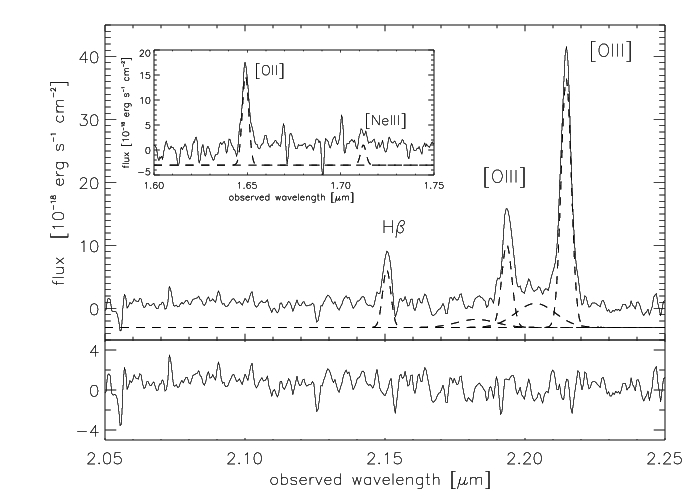}
\caption{Integrated spectrum of SW022513 showing H$\beta$ and the
[OIII]$\lambda\lambda$4959,5007 doublet. Note the broad [OIII]
components. Hatched lines show Gaussian fits with parameters given in
Table~\ref{tab:spectraSW022513} and are shifted by an arbitrary value
along the ordinate. The lower panel shows the fit residuals. The inset
shows [OII]$\lambda\lambda$3726, 3729 (the doublet is blended due to
the broad line widths and is fitted with a single Gaussian profile for
simplicity) and [NeIII]$\lambda$3869 in the H-band part of the
spectrum.}
\label{fig:SW022513_intspec}
\end{figure}

We show the integrated spectrum of SW022513 in
Figure~\ref{fig:SW022513_intspec}. Line properties are broadly
consistent with those found by \citet{polletta08} with ISAAC longslit
spectroscopy at 3$\times$ lower spectral resolution and lower
signal-to-noise ratios.  The lines are spectrally well resolved with
typical widths of FWHM$\ge$700 km s$^{-1}$. Due to the broad width of
the lines we did not resolve the individual components of the
[OII]$\lambda\lambda$3726,3729 doublet.

SW022513 has luminous [OII]$\lambda$3727, [OIII]$\lambda\lambda$4959,
5007, and H$\beta$ line emission, with [OIII]$\lambda$5007/H$\beta =
6$. [NeV]$\lambda\lambda$3346,3426 fall outside the atmospheric
windows, but we detect [NeIII]$\lambda$3869. We will argue in
\S\ref{ssec:localizing} that, for galaxies with the overall
characteristics of SW022513, the [OIII]/H$\beta$ ratio and the
detection of [NeIII] suggest that most of the warm ionized gas is
photoionized by the AGN.

\begin{figure}
\centering
\includegraphics[width=0.7\textwidth]{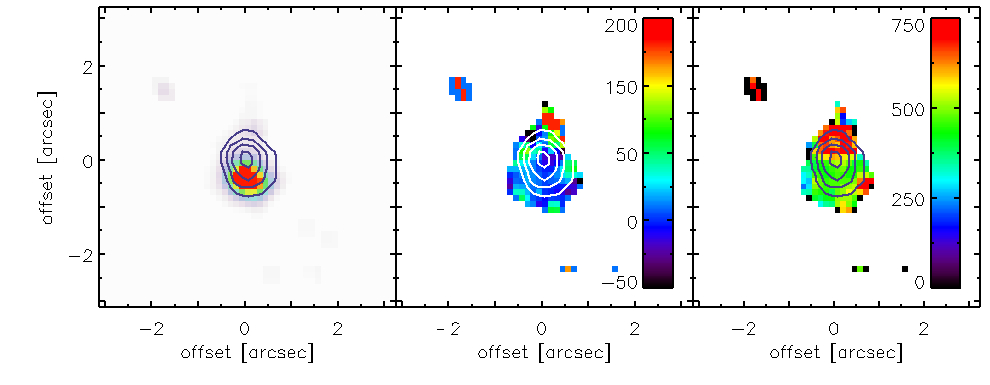}\\
\includegraphics[width=0.7\textwidth]{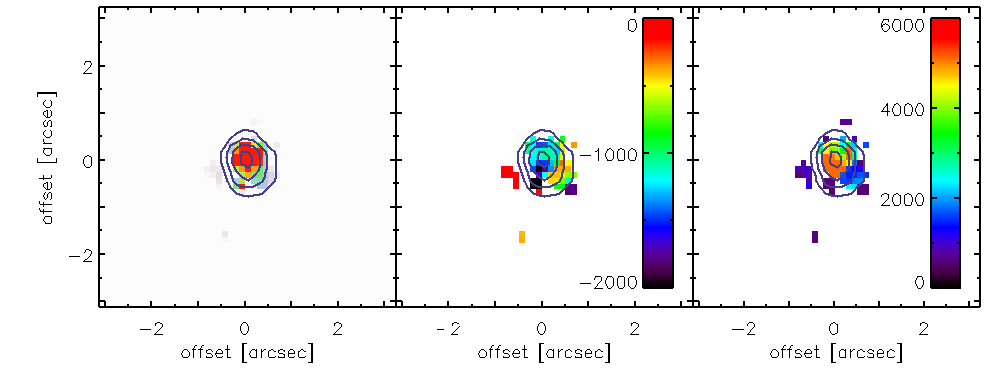}\\
\caption{[OIII]$\lambda$5007 emission-line maps for SW022513 for the
narrow (top panel) and broad component (bottom panel). From left to
right, we show the maps of integrated fluxes, velocities relative
to the velocity at the positio of the continuum peak, and
line widths. Color bars indicate the velocities and line widths in km
s$^{-1}$. Contours show the continuum morphology in all panels.}
\label{fig:SW022513_maps}
\end{figure}

Line emission in SW022513 is extended over sizes of
1.6\arcsec$\times$2.5\arcsec\ in right ascension and declination,
respectively. The [OIII]$\lambda$5007 emission-line morphology is
shown in the upper left panel of Figure~\ref{fig:SW022513_maps}. The
[OIII]$\lambda$5007 line flux does not peak on the continuum peak, but
is offset by 0.75\arcsec\ to the South (corresponding to a projected
distance of about 5 kpc).

Emission-line profiles are complex. We identify broad, blueshifted
components in [OIII]$\lambda\lambda$4959,5007 and H$\beta$. Unlike
[NeIII]$\lambda$3869 and [OII]$\lambda\lambda$3726,3729, the H$\beta$
and [OIII]$\lambda$5007 lines are fairly bright and do not suffer
blending of several components with uncertain relative line
widths. They are therefore particularly suited to investigate their
line profiles, and we will focus our discussion on those two
lines. Line profile fits to [OIII]$\lambda$4959 yield similar results
as [OIII]$\lambda$5007 within the measurement uncertainties. 

The broadest [OIII]$\lambda$5007 component is found near the continuum
peak, with FWHM$=$5078 km s$^{-1}$ and an offset of $-$1314 km s$^{-1}$
relative to the narrow [OIII]$\lambda$5007 component measured in the
same spectrum. At the same position we also detect a broad component
in H$\beta$, with FWHM$=$1000 km s$^{-1}$ and a blueshift of
-183~km~s$^{-1}$ relative to the narrow component (which has FWHM=369
km~s$^{-1}$). All emission-line properties and their uncertainties are
listed in Tables~\ref{tab:spectraSW022513_contpeak},
\ref{tab:spectraSW022513_OIIIpeak}, and
\ref{tab:spectraSW022513_north}.

We obtained maps of the relative velocities and line widths from
fitting emission lines extracted from apertures of 3 pixels $\times$3
pixels (0.4\arcsec$\times$0.4\arcsec) at all positions where
[OIII]$\lambda$5007 line emission was detected at $>$3$\sigma$,
adopting the prodedure outlined in \citet{nesvadba08b}.  To account
for the complex profile of the line, we performed fits with 2 Gaussian
components for each line. We fitted the
[OIII]$\lambda\lambda$4959,5007 doublet simultaneously, requiring that
both lines have the same redshift and line width, and a ratio as
expected between the two components of R(5007,4959)$=$3. Overall this
gives a line fit of 4 Gaussian components, where the narrow and broad
components of [OIII]$\lambda$4959 and [OIII]$\lambda$5007 are required
to have the same kinematics and a line ratio consistent with their
Einstein coefficients.

The maps of relative velocities and line widths (full width at half
maximum) of the narrow [OIII]$\lambda$5007 component are shown in the
upper middle and right-hand panel of Figure~\ref{fig:SW022513_maps},
respectively. Velocities are relatively uniform across most of the
source, with a sudden redshift jump of $\sim$200 km s$^{-1}$ northward
of the nucleus. The corresponding maps for the broad component are
shown in the lower panel of Figure~\ref{fig:SW022513_maps}. 
We detect a very broad (FWHM$\sim$5000 km s$^{-1}$) component of
[OIII]$\lambda\lambda$4959,5007 associated with the continuum peak,
and a somewhat less extreme, but nonetheless broad component with
FWHM$\sim$1000 km s$^{-1}$ associated with the peak of the
[OIII]$\lambda\lambda$4959,5007 emission. 

To map line ratios relative to [OIII]$\lambda$5007 we also fitted the
spatially resolved emission of H$\beta$ and
[OII]$\lambda\lambda$3726,3729, where we assumed a single Gaussian
distribution for the [OII] doublet which is not spectrally
resolved. In Figure~\ref{fig:SW022513_ratiomaps} we show the maps of
line ratios relative to the narrow [OIII]$\lambda$5007 component. The
ratios of [OII]$\lambda$3727 and of H$\beta$ to [OIII]$\lambda$5007
are smallest near the [OIII]$\lambda$5007 emission-line peak, and have
larger values near the brightest continuum emission. 

\begin{figure} \centering
\includegraphics[width=0.4\textwidth]{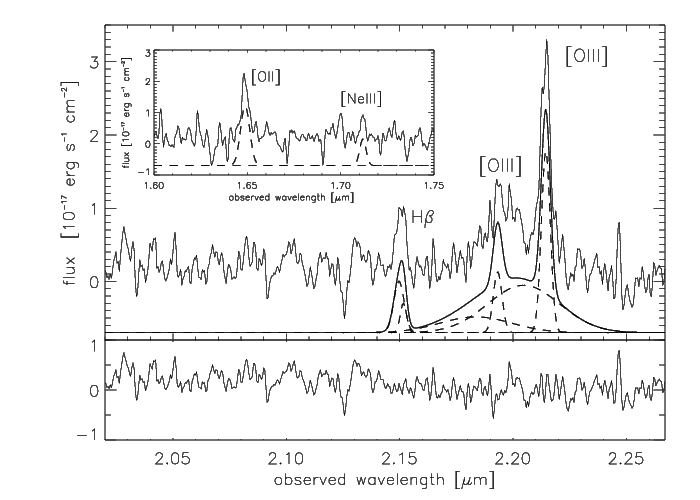}\\
\includegraphics[width=0.4\textwidth]{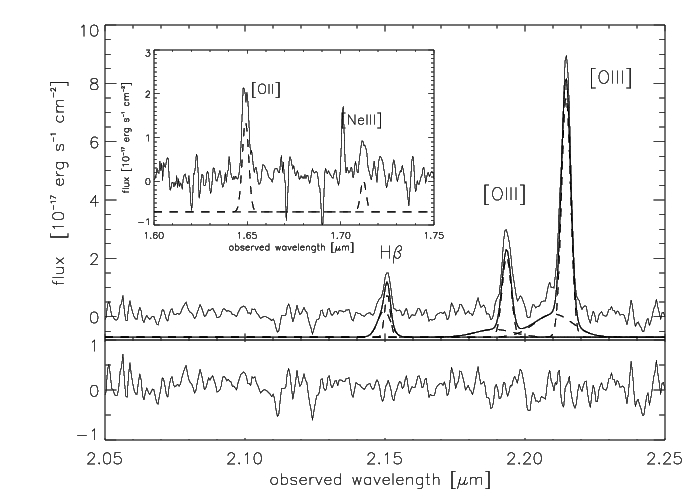}\\
\includegraphics[width=0.4\textwidth]{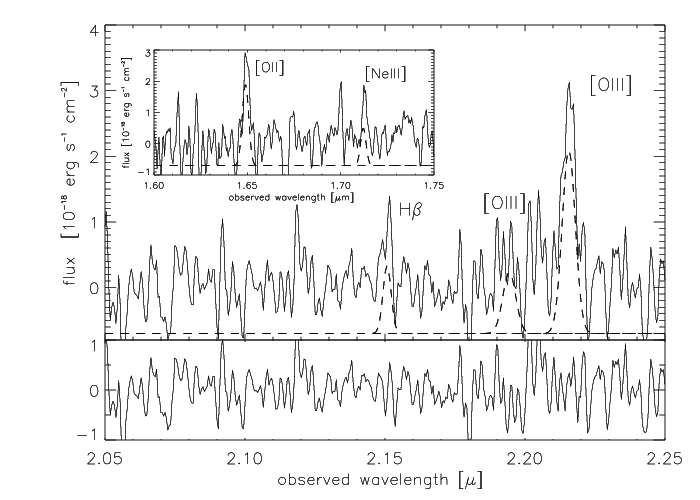}\\
\caption{Spectra of SW022513 extracted from
0.4\arcsec$\times$0.4\arcsec\ box apertures centered on the continuum
peak (upper panel), the peak in [OIII]$\lambda$5007 emission-line
surface brightness (mid panel). The lower panel shows the integrated
spectrum of the faint emission-line region north of the nucleus
(extracted from an aperture of 0.4\arcsec$\times$0.5\arcsec\ in right
ascension and declination, respectively).}
\label{fig:SW022513_spectra}
\end{figure}

H$\beta$ and [OII]$\lambda\lambda$3726,3729 are too faint to map
the kinematics of multiple components. Therefore we extracted
integrated spectra associated with the peaks in
[OIII]$\lambda\lambda$4959,5007 and continuum emission respectively,
and from the faint, extended region to the North. These spectra are
shown in Figure~\ref{fig:SW022513_spectra} and their properties are
summarized in Tables~\ref{tab:spectraSW022513_contpeak},
\ref{tab:spectraSW022513_OIIIpeak}, and
\ref{tab:spectraSW022513_north}.

\subsubsection{SW022550}
SW022550 is a luminous line emitter in the rest-frame UV as discussed
by \citet{polletta08}. Unfortunately at z$=$3.867 it is at a somewhat
unfavourable redshift for ground-based observations, where
[OIII]$\lambda$5007 and [OII]$\lambda$3726,3729 fall outside the
atmospheric windows. However, we did identify the [OIII]$\lambda$4959
line at $\lambda$=24182 \AA, corresponding to a redshift $z=3.876$,
which is consistent with the redshifts measured in the rest-frame UV.
The [OIII]$\lambda$4959 line is very broad, FWHM=2212 km s$^{-1}$, not
very different from the width of the rest-frame UV lines, which have
line widths of up to few 1000 km s$^{-1}$ \citep{polletta08}. The line
flux is F(4959)$=$8.2$\times 10^{-16}$ erg s$^{-1}$ cm$^{-2}$. For
F(4959)/F(5007) = 0.3 this corresponds to a [OIII]$\lambda$5007 flux
of F(5007)$=$2.5$\times 10^{-15}$ erg s$^{-1}$ cm$^{-2}$. We did not
detect H$\beta$ but place an upper limit of $4.3\times 10^{-16}$ erg
s$^{-1}$ cm$^{-2}$ at 3$\sigma$ significance, assuming it has a
similar width as [OIII]$\lambda$4959.

We also detect the [NeV]$\lambda\lambda$3346,3426 doublet, although
[NeV]$\lambda$3346 is heavily blended with night sky line residuals,
making it difficult to measure anything but the line core. These lines
give redundant kinematic and flux information and are well fit
with a common redshift of z=3.861 and line width of FWHM$=$2985 km
s$^{-1}$. For [NeV]$\lambda$3426 we measure a flux of $4.3\times
10^{-16}$ erg s$^{-1}$, and the core of [NeV]$\lambda$3346 is
consistent with being 3$\times$ fainter as expected from the
transition probabilities of the two lines. The rest-frame optical
spectral properties of SW022550 are listed in
Table~\ref{tab:spectraSW022550}.

All line emission appears compact and associated with the compact
continuum emission.

\section{Two quasars with giant narrow-line regions at z$\ge$3.5}
\label{ssec:localizing}

Our spectroscopic results suggest that the AGN is the dominant
source of ionization in our targets. For SW022550 this follows
directly from the detection of [NeV]$\lambda$3426. Line emission from
this galaxy is not spatially resolved, the size of the seeing disk of
0.6\arcsec\ implies an upper limit on the radius of the NLR of about
2.5 kpc, provided that we did not miss fainter, extended structures
due to the somewhat unfavorable redshift of the source
(\S\ref{ssec:emissionlinegas}).

At the somewhat lower redshift of SW022513, [NeV]$\lambda$3426 does
not fall into the atmospheric windows. We do however, observe
[NeIII]$\lambda$3869 and find a relatively high [OIII]/H$\beta$ ratio
of 6. A priori, both features could be produced either by the AGN or
by intense star formation in galaxies with high ionization parameters
and low metalicities, leading to hot electron temperatures, like, e.g.,
in high-redshift HII or Lyman-break galaxies \citep[e.g.,][]{pettini01,
fosbury03, nesvadba06a, nesvadba07a, nesvadba08a}. \citet{villar08}
find that star formation may photoionize parts of the gas in 3/50
type-2 quasars drawn from the SDSS; however, these are much less
powerful than the targets we observe, and have less luminous optical
lines, e.g., ${\cal L}([OIII])\sim 10^{42}$ erg s$^{-1}$ compared to
few$\times 10^{44}$ erg s$^{-1}$ for SW022513
(Table~\ref{tab:spectraSW022513}). We are not aware of any
star-forming galaxy without AGN and with ${\cal L}([OIII])=10^{44}$
erg s$^{-1}$.

In addition, the bright 24$\mu$m and millimeter continuum
emission of SW022513 and SW022550 shows these are dusty galaxies. Dust
and metal lines are very efficient coolants, so that for a given
ionizing spectrum (e.g., due to intense star formation), we may
expect lower electron temperatures than in HII galaxies and LBGs,
which would lead to lower [NeIII] fluxes and lower [OIII]/H$\beta$
ratios. The FIR/millimeter properties of SW022513 and SW022550 are
indistinguishable from those of submillimeter-selected starburst
galaxies at z$\ge$2, which have supersolar metallicities
\citep{tecza04, swinbank04, nesvadba07a} and fall near the local
mass-metallicity relationship of \citet{tremonti04}. In the absence of
an AGN, these galaxies have [OIII]/H$\beta \le 1$
\citep{takata06,nesvadba07a}, and [NeIII] is generally not observed.
For SW022550 and SW022513, the mass-metallicity relationship would
imply a gas-phase oxygen abundance of $12+log(O/H)\sim 9.1$,
supersolar even when accounting for the (large) scatter in the
relationship (and supersolar by 0.4 dex when taken at face
value). \citet{polletta08} came to a similar conclusion from the
rest-frame UV line ratios of SW022550. Roughly solar metallicities
have also been found by \citet{humphrey08} for dusty, massive radio
galaxies at high redshift.

The line emission in SW022513 is well resolved spatially. This implies
that the QSO narrow-line region (NLR) extends to radii of at least
1.25\arcsec\ (corresponding to $\sim$10 kpc). Most notably, the peak
of the narrow [OIII]$\lambda\lambda$4959,5007 line is offset by
0.75\arcsec\ (5 kpc) towards South from the continuum peak and the
peak of broad [OIII]$\lambda\lambda$4959,5007 emission. A 
spatial offset between broad and narrow line emission has previously
been reported for the submillimeter-selected quasar SMMJ1237$+$6203 at
z=2.1 \citep{alexander10}. Without additional constraints, this offset
could be interpreted as a signature of an outflow of turbulent gas, at
several 10s of kpc from the galaxy traced by the narrow-line emitting
gas.

Our data on SW022513 allow us to perform a more complete analysis 
which suggests a different scenario.  First, the
broad [OIII] line emission is spatially coincident with the continuum
peak, implying that the emission arises in the nuclear region of the
galaxy and thus near the AGN. We also find that near the peak of
narrow [OIII] emission, the [OII]/[OIII] and the H$\beta$/[OIII] line
ratios are smaller than near the continuum peak, and hence near the peak
of broad [OIII] and the likely location of the AGN (Figure 5).  This
implies that the narrow line region is more highly ionized than the
emission immediately surrounding the AGN.  This increase can be caused
by a harder radiation field for a constant ionization parameter (ratio
of radiation field intensity to gas density), an increasing ionization
parameter, or a decrease in gas-phase metallicity as we move from the
continuum to the narrow-line peak. It could be a combination of all
three \citep[see][for a more detailed analysis in related
environments]{humphrey08}.

Given the bright FIR luminosity of SW022513, the extinction in the
circum-nuclear region is likely to be high.  However, large-scale
extinction of the emission line gas cannot solely determine the
variation in [OIII]/[OII] line ratio. The same trend is also found in
[OIII]/H$\beta$ and given their similar wavelengths (including the [OIII]
line at 4959\AA) the influence of extinction must be relatively small.
In relation to the nebula itself, the effect of dust on the ionization of
the gas is complex. Since dust provides an additional source of cooling
and competes directly with the gas for ionizing photons due to its large
cross section over a wide range of wavelengths but also depletes metals
from the gas (some of which provide important cooling lines like Oxygen
and Carbon), a dusty gas will have a different ionization structure
depending on the gas-to-dust ratio and amount of metal depletion.
So it is not entirely clear what its impact will be, but it could well
increase the ionization of the gas. 

All these scenarios would produce the situation we observe and suggest
that the characteristics of the extended ionized gas in SW022513 are
consistent with a narrow-line region surrounding a luminous obscured
quasar and extending over a radius of $\sim$10 kpc. Radii of few 10s
of kpc are expected for the narrow-line regions of the most powerful
quasars \citep{netzer04} and consistent with empirical constraints for
optically selected quasars \citep{bennert02} and also type-2 quasars
\citep{greene11}. In addition, the asymmetry of the narrow [OIII] line
morphology relative to the nucleus supports this scenario as
narrow-line regions are often highly elongated and asymmetric. Given
the bright FIR luminosity of our targets of $10^{13} {\cal L}_{\odot}$
\citep{polletta08}, we expect these are highly dust-enshrouded, and
that most of the rest-frame UV/optical light is obscured on kpc
scales, with ionizing and non-ionizing radiation escaping to larger
distances only along relatively few lines of sight that are
comparably free of dust. This is analogous to the situation observed
for local Seyfert 2s in their extended narrow-line regions
\citep[e.g.,][see \citealt{lehnert09} for an example at z$>$2.]{veilleux03}.

This does not imply that the quasar also dominates
the gas kinematics. Quasar illumination cones without strong
mechanical effect are known at low redshift \citep{veilleux03} as well
as at z$=$2 \citep{lehnert09}. Note that this applies to radio-quiet
quasars. Radio-loud AGN, including radio galaxies, often have much
more extended emission-line regions with very energetic emission-line
kinematics \citep[e.g.][]{mccarthy96, villar99, baum00, villar03,
nesvadba06b, nesvadba08b}.

\subsection{Ionized gas mass} 
\label{ssec:masses}

The molecular gas strongly outweights the ionized gas in strongly
star-forming mergers at z$\sim$2 \citep{nesvadba07a}, but not in
powerful radio galaxies at similar redshifts \citep[][Nesvadba
et al., in prep.]{nesvadba08b} where much of the interstellar medium
appears photoionized by the AGN
\citep[e.g.,][]{villar97,villar03,humphrey08} and accelerated by the
radio source \citep[e.g.,][]{nesvadba08b}. We will now use the
H$\beta$ emission-line flux measured in SW022513 to illustrate that
the ionized gas mass in this obscured quasar could plausibly be as
large as the molecular gas mass.

Similar to \citet{nesvadba06b,nesvadba08b} we will assume case B
recombination to estimate the ionized gas mass from the luminosity of
the Balmer lines, setting
\begin{eqnarray} {\cal M}_{H\beta}\ = 28.2\times 10^8\ {\cal
L}_{H\beta,43}\ n_{e,100}^{-1}\ M_{\odot},
\end{eqnarray} where ${\cal L}_{H\beta,43}$ is the H$\beta$ luminosity
in units of $10^{43}$ erg s$^{-1}$, and $n_{e,100}^{-1}$ is the
electron density in units of 100 cm$^{-3}$. This relationship is
equivalent to Equation (1) of \citet{nesvadba06b} assuming a Balmer
decrement H$\alpha$/H$\beta =$2.9, and follows directly from \citet{osterbrock89}.

This estimate has two major uncertainties. First, the
[OII]$\lambda\lambda$3726,3729 line doublet is blended, so we have no
direct estimate of the electron density. Second, from H$\beta$ alone
we cannot measure the extinction. We can loosely constrain these
values with rather generic arguments. First, the presence of luminous
forbidden emission lines like the [OII]$\lambda\lambda$3726,3729
doublet suggests that electron densities are below the critical
densities for collisional deexcitation, a few 1000 cm$^{-3}$, and we
will in the following adopt a fiducial value of 1000 cm$^{-3}$
\citep[radio galaxies and submillimeter galaxies at similar redshifts
have electron densities of few 100
cm$^{-3}$][]{nesvadba06b,nesvadba07b,nesvadba08b}. Second, we
will assume an average extinction of about $A_V\sim 2$ mag, which is
the average in submillimeter-selected galaxies without strong AGN
\citep{smail04} that are the closest analogs to our galaxies in the
far-infrared/submillimeter \citep[and consistent with a nuclear
A$_V\sim$4.6 mag derived
by][]{polletta08}. \citet{nesvadba08b} find A$_V=$1$-$4 mag from the
H$\alpha$/H$\beta$ line ratios in the extended ionized gas of z$\sim$2
radio galaxies. For SW022513, this gives an H$\beta$ luminosity of ${\cal
L}_{H\beta} = 7\sim \times 10^{44}$ erg s$^{-1}$, a factor $\sim 10$
higher than the observed value (Table~\ref{tab:spectraSW022513}).

With these two assumptions, we find an ionized gas mass of $2\times
10^{10}$ M$_{\odot}$ (compared to $2\times10^{9}$ M$_{\odot}$ if we
strictly use the observed H$\beta$ flux). This can obviously only be
an order-of mass estimate, but illustrates that the mass of ionized
gas in SW022513 is likely between 10\% and 100\% of the molecular gas
mass. This is in the same range as found for radio galaxies, and a
factor 100$-$1000 higher than for strongly star-forming galaxies. For
SW022550 we only have an upper limit on the H$\beta$ flux, which is
not very constraining. The above considerations suggest that the
ionized gas mass estimate and derived quantities can only be accurate
at an order-of-magnitude level.

\subsection{What is the systemic redshift?}
\label{ssec:systemicredshift}
The lack of robust measurements of the systemic redshift is a major
uncertainty in all studies of the gas dynamics in galaxies at high
redshift, and yet, is indispensible to interpret the emission-line
kinematics. At distances where direct measurements of stellar
absorption line kinematics in AGN hosts are possible, observations
suggest that the narrow emission-line components show only moderate
offsets from the systemic velocity. This includes Seyfert galaxies
\citep{nelson96} as well as powerful, type-2 quasars in the SDSS
\citep{greene05}, and also galaxies with very broad blueshifted
[OIII]$\lambda\lambda$4959,5007 components \citep[][]{wong06}.

The narrow [OIII]$\lambda$5007 component in SW022513 has a relatively
small velocity gradient, $\Delta v \sim$200 km s$^{-1}$
(Figure~\ref{fig:SW022513_maps}), and lines of different species
of ionized gas have very similar velocities. This is broadly
consistent with what would be expected for rotation or merger-driven
kinematics of galaxies with stellar masses of $\sim 10^{11} M_{\odot}$
and is in the velocity range of submillimeter galaxies at z$\ge$2
\citep[][]{swinbank06,nesvadba07a} as well as nearby ULIRGs
\citep{colina05}. We will thus in the following assume that the narrow
optical emission lines of SW022513 trace the systemic redshift within
$\le 100$ km s$^{-1}$. Specifically we will use the redshift of the
narrow H$\beta$ component near the continuum peak (which we identify
as the nucleus, \S\ref{ssec:localizing}), z$=$3.4247
(Table~\ref{tab:spectraSW022513_contpeak}) to approximate the systemic
redshift. H$\beta$ is measured at good signal-to-noise
(SNR$=$8.5$\sigma$) and is more representative for the overall gas
kinematics than [OIII]$\lambda\lambda$4959,5007, which are very
sensitive to ionization effects.

\subsection{Blue wings as signatures of outflows} 
\label{ssec:bluwings}
If the narrow lines of AGN host galaxies are approximate tracers of
the systemic velocity, then the broad, blueshifted components often
found in [OIII]$\lambda\lambda$4959,5007 as well as other lines, most
likely trace gas that is in outflow
\citep[e.g.,][]{heckman81,greene05b, morganti05, nesvadba07b,
komossa08, holt08, nesvadba08b, greene09, spoon09a, spoon09b,
nesvadba10a}. Scalings between the [OIII]$\lambda$5007 width and AGN
power \citep[either radio power;][or bolometric luminosity;
\citealt{greene05b}]{heckman81} broadly support this
picture. Following these previous analyses, we therefore consider the
broad wings of [OIII] and other lines in SW022513 and the broad [OIII]
lines in SW022550 as signatures of outflows. 

In SW022513 the broadest line widths of FWHM$\ge$5000 km s$^{-1}$ seen
in [OIII] are broader than the other lines in the same aperture, and
broader than all other lines including [OIII] in all other
apertures. The [OIII] emissivity is very sensitive to electron
temperature and excitation conditions, but not the total gas mass, so
this component is unlikely to trace large amounts of gas. The most
strongly blueshifted gas seen in the more representative H$\beta$
line, blueshifted by of order $-$1000 km s$^{-1}$ relative to
systemic, likely gives a more robust estimate of the overall gas
kinematics (see also \S\ref{ssec:molgas}). We ran a Monte-Carlo
simulation to infer for what H$\beta$/[OIII]$\lambda$5007 line ratio
we could have detected a broad H$\beta$ component with FWHM=5000 km
s$^{-1}$ and at the same redshift as the broad
[OIII]$\lambda\lambda$4959,5007 lines \citep[see \S~3.1 of ][for more
details]{nesvadba10b}. For 1000 throws we found that H$\beta$ would
have been detected at 3$\sigma$ for [OIII]$\lambda$5007/ H$\beta \le
6.6$, compared to [OIII]$\lambda$5007/H$\beta$=3.8 for the narrow
component in the same spectrum. If the assumptions of
\S\ref{ssec:masses} hold, then this would imply a $3\sigma$ upper
limit to the mass of high-velocity ionized gas (with FWHM$=$5000 km
s$^{-1}$) of $1.5\times 10^{8} M_{\odot}$ (neglecting extinction), or
$1.5\times 10^9$ M$_{\odot}$ (for A$_V=$2 mag; see
\S\ref{ssec:masses}), roughly 10\% of the mass found at more moderate
velocities. Interestingly, the higher [OIII]$/$H$\beta$ ratio suggests
this gas is more highly ionized than the narrow-line emitting gas.

Notwithstanding, H$\beta$ also shows the most strongly blueshifted
components near the nucleus (see Tables
\ref{tab:spectraSW022513_contpeak} to
\ref{tab:spectraSW022513_north}). This is expected if the gas flows
are driven by the AGN and decelerate as they interact with ambient gas
at larger radii.

\subsection{Comparison with molecular gas} 
\label{ssec:molgas}

\citet{polletta10} recently discussed integrated CO(4--3)
emission-line spectroscopy of SW022513 and SW022550 obtained with the
IRAM Plateau de Bure Inteferometer at a spatial resolution of
$\ge$4\arcsec.  Both galaxies have luminous millimeter CO(4--3) line
emission that corresponds to $\sim 4\times 10^{10}$ M$_{\odot}$ of
molecular gas, assuming a factor of 0.8 M$_{\odot}$ K km s$^{-1}$
pc$^2$ to convert CO luminosity to a molecular gas mass. SW022550 has
a double peaked profile with two components separated by 440 km
s$^{-1}$ in velocity.
 SW022513 has a single, very broad component with FWHM$\sim$1000
km s$^{-1}$, which is not very well fitted with a single Gaussian.

In Figure~\ref{fig:SW022550_CO_Hbeta} we compare the CO(4--3) and
[OIII]$\lambda$4959 line profiles of SW022550. The redshift of [OIII],
z$=$3.876$\pm$0.001 falls between that of the two CO peaks.
In SW022513, the CO(4--3) line
profile shows an excellent match with the H$\beta$ profile extracted
from the nuclear aperture centered on the continuum peak
(Figure~\ref{fig:SW022513_CO_Hbeta}). This includes in particular the
broad, blueshifted wings with velocities of up to -1000 km s$^{-1}$
relative to the narrow H$\beta$ component. As we argued in
\S\ref{ssec:systemicredshift}, the narrow H$\beta$ component most
likely approximates the systemic velocity to about 100 km s$^{-1}$.

Comparison with telluric night-sky lines suggests that
the absolute wavelength scale in our SINFONI data is accurate to about
20 km s$^{-1}$. In the millimeter, wavelengths are measured relative
to a local oscillator, therefore uncertainties in wavelength
calibration are much smaller and can be neglected.

Formally, our Gaussian fits imply a velocity offset of -183$\pm$67 km
s$^{-1}$ for an assumed Gaussian line core of the blueshifted
H$\beta$ component, and of -181$\pm$47 km s$^{-1}$ for the Gaussian
core of the CO(4-3) line. Molecular line emission is therefore found
with blueshifts of up to $-$900 to $-$1000 km s$^{-1}$ relative to the
narrow H$\beta$ component, which, as argued in
\S\ref{ssec:systemicredshift}, is our most robust measure of the
systemic redshift. Regardless of the detailed match between molecular
and ionized gas, and neglecting possible projection effects, this is
likely to be more than the escape velocity (\S\ref{ssec:willthegasescape}).

\section{The wind in a z$=$3.5 quasar}
\label{sec:wind}

Our targets have all the hallmarks of being in a short, decisive, and
very complex stage of their evolution, where star formation, AGN, and
gravitational interactions are likely to release an energy of
$10^{59-60}$ erg, the equivalent of the binding energy of the host
galaxy and its dark-matter halo. At their current luminosities, the
AGN in SW022513 and SW022550 may release such energies in a
few $10^{6-7}$ yrs, which corresponds to the lifetimes over which AGN
may maintain bolometric luminosities $>10^{46}$ erg s$^{-1}$
\citep{hopkins05}. Similarly, the canonical energy release of
$10^{51}$ erg per supernova explosion corresponds to a total energy of
about $\times 10^{58}$ erg released in the formation of $10^{11}$
M$_{\odot}$ of stars, and observations have shown that the hydrostatic
pressure of intense starbursts may balance gravity in z$=$2 galaxies
very similar to our targets \citep{nesvadba07a}. Major mergers
naturally release the equivalent of the binding energy of the merger
remnant during the interaction. Hence, each of these mechanisms could
have a strong influence on the kinematics and thermal state of the gas
in our targets. What mechanism is dominating the gas dynamics in
SW022550 and SW022513?

Important to address this question is not only the total energy
injection, but also the timescales over which this energy is released,
and the efficiency with which the energy output is turned into an
input of (thermal or mechanical) energy into the ambient
gas. \citet{narayanan08} find in SPH simulations of a gas-rich, major
merger associated with intense star formation and AGN activity, that
the AGN affects the gas more drastically than star formation and
gravitational collapse because of the shorter energy injection
time. Star formation and gravitational interaction release their
energy in few $10^{8}$ yrs and up to $10^9$ yrs, respectively, much
longer than the $10^{6-7}$ yrs lifetime of the AGN.

Observations lead to a similar conclusion. The spectral signatures of
winds in submillimeter galaxies without strong AGN component appear
much more subtle, with smaller blueshifts, and a smaller mass of
entrained material \citep{nesvadba07b}. Overall, these galaxies have
velocity gradients and line widths of up to a few 100 km s$^{-1}$, and
more regular line profiles \citep[e.g.,][]{swinbank05, swinbank06,
takata06} in spite of stellar masses and star-formation rates very
similar to our targets. Low-redshift ULIRGs including very advanced
mergers have velocity gradients and line widths in a similar
range \citep{colina05}, and are much smaller than $\sim$1000 km
s$^{-1}$ at any stage of merging.

SW022513 and SW022550 both show signatures that the AGN does affect
the warm ionized ISM, evidenced through the broad, blue components in
SW022513 and the very broad FWHM of [OIII]$\lambda$4959 in SW022550
(FWHM$>$2000 km s$^{-1}$). However, FWHM$>$1000 km s$^{-1}$ are only
found around the nucleus in both cases, and , since we cannot
constrain extinction very well, it is not clear if the warm ionized
gas represents a major fraction of the ISM (\S\ref{ssec:masses}). The
CO profiles of both galaxies, which trace about $4\times 10^{10}$
M$_{\odot}$ in molecular gas in each source, are very different in
SW022513 and SW022550 \citep{polletta10}. In SW022550 the CO profile
resembles the fairly common double-peaked profiles found, e.g, in
submillimeter galaxies, where they are commonly interpreted as
signatures of mergers or rotating disks \citep[and where even disks
may be signatures of (advanced) mergers;][]{downes98}, in agreement
with the merger models of \citet{narayanan08}. In either case, the
barycenter falls roughly inbetween the two CO peaks. This could
suggest that the AGN affects only a small part of the multiphase ISM
in SW022550, traced by the broad [OIII] line.

SW022513 is however different. The CO line profile is irregular and
neither a clear single nor double-peaked Gaussian. The line extends to
large relative velocities of up to $-$1000 km s$^{-1}$ from systemic
(\S\ref{ssec:molgas}), similar to the ionized gas near the AGN
(Fig.~\ref{fig:SW022513_CO_Hbeta}). Outflows driven by the AGN are the
most common interpretation of blue, broad wings of {\it ionized} gas
(\S\ref{ssec:bluwings}), which may suggest that parts of the {\it
molecular} gas in SW022513 may be tracing outflowing gas as well.
Winds are inherently multi-phase phenomena, where the warm ionized gas
seen through optical line emission is being entrained by a hot,
tenuous medium. It may therefore {\it a priori} not be entirely surprising 
to find an associated phase of molecular gas. However, this does raise
fundamental questions of how molecular gas, which is 1-2 orders of
magnitude denser ($N\sim 10^{3-4}$ cm$^{-3}$) than ionized gas ($N\sim
10^{2-3}$ cm$^{-3}$) is accelerated to high velocities. This could be
alleviated if much of the gas is diffuse and distributed across the
galaxy, perhaps through tidal effects or a starburst-driven wind
\citep{narayanan08}. 

It is also possible that the molecular gas is
forming {\it in situ} in the outflow, as suggested by recent studies
of the detailed mass and energy exchange in turbulent, multiphase gas
in nearby extragalactic environments with galaxy-wide shocks,
including radio galaxies \citep[][see also Krause \& Alexander, 2007
]{guillard09, nesvadba10a, ogle10}. In these
scenarios, the ionized line emission may arise from the turbulent
mixing interfaces associated with the same clouds, which would explain
why ionized and molecular gas have similar velocities.
\citet[][]{papadopoulos10, papadopoulos08} recently found bright
CO(6-5) and CO(3-2) line emission in the nearby radio galaxy 3C293,
which is a posterchild of a jet-driven outflow \citep{morganti05,
emonts05}. CO line ratios in 3C293 suggest that most of the molecular
gas is dense, turbulent, and gravitationally unbound, as expected in
a multiphase scenario.

Possible kinematic signatures of outflows from star-forming AGN host
galaxies traced through CO line emission have previously been reported
in a few nearby cases \citep{appleton02, sakamoto06, alatalo10,
iono07, feruglio10, irwin10}. These galaxies have AGN power and
star-formation rates that are lower by 1-2 orders of magnitude than in
SW022513. Outflow velocities are typically a few 100 km s$^{-1}$, and entrainment
rates can be up to few 100 M$_{\odot}$ yr$^{-1}$. The CO profiles
found by \citet{sakamoto06,alatalo10} are well matched by HI
absorption-line profiles tracing neutral outflows, and providing
robust evidence for the outflow interpretation.  However, typically
these winds include only a small fraction of the CO luminosity, unlike
in SW022513 where most of the CO emission appears to be
blueshifted. This may partially be due to the higher-excitation gas
probed in the J$=$4--3 transition, or it may imply that more gas is
being entrained compared to low-redshift galaxies. This is not
implausible, given the much greater AGN power in SW022513 compared to
nearby Seyfert galaxies, and the large molecular gas mass. We will in
the following quantify the necessary and available amounts of kinetic
energy in SW022513 to investigate whether the AGN may plausibly drive
the gas to the velocities observed.

\subsection{Kinetic energy of the gas}
\label{ssec:ekingas}

In Figure~\ref{fig:SW022513_CO_Hbeta} we compared the line profiles of
H$\beta$ and CO, and argued that, by analogy with a large number of
previous AGN studies, the systemic velocity is typically well
approximated by the narrow component of optical emission lines,
whereas the outflowing component is in the blue wing.  To estimate the
kinetic energy in the outflowing gas, we therefore decompose the CO
and H$\beta$ line profiles into a 'systemic' and a blueshifted
component, finding that about 40\% of the total CO(4-3) emission-line
luminosity is in the outflowing component (and about 30\% of
H$\beta$). Assuming that the CO-to-H$_2$ conversion factor and the CO
excitation are independent of velocity, this suggests that 40\% of the
molecular gas is in the blueshifted component.

Based on these assumptions, we estimate the kinetic energy directly
from the line profile, by setting
\begin{eqnarray} E_{kin} = 1/2\ \sum_{v=0}^{v_{max,blue}}M_i\ v^2_i,
\end{eqnarray} where $M_i$ is the gas mass in each velocity bin $v_i$
(relative to the systemic velocity, \S\ref{ssec:systemicredshift}),
and including only the flux from the blueshifted component. This
corresponds to a kinetic energy of about $4\times 10^{58}$ erg. Adding
the ionized gas mass would add another $0.2-2\times 10^{58}$ erg,
depending on extinction (see \S\ref{ssec:masses}). Obviously, these
estimates have large systematic uncertainties related to the molecular
gas mass estimate, excitation conditions, and velocity estimates
including projection effects. We therefore consider this estimate
accurate to about an order of magnitude.

\subsection{Is the AGN capable of driving the outflow?}

At the beginning of the section we have argued that a starburst-driven
wind and an interaction are unlikely to accelerate significant amounts
of gas to velocities of $-$1000 km s$^{-1}$. We will now investigate
for two popular AGN feedback mechanisms, if they may plausibly explain
the observed gas kinematics.

\subsubsection{Radiation pressure from AGN and starburst} 
\label{ssec:radiationpressure}

Radiation pressure is often invoked to
explain how AGN may expel significant
amounts of gas from their host galaxies
\citep[e.g.,][]{king03,murray05}. 
\citet{murray05} discussed the outflow velocities that can be produced
by radiation pressure from AGN and intense starbursts, finding that
(their Equation~17)
\begin{eqnarray} V(r) = 2\sigma \sqrt{ (\frac{{\cal L}}{{\cal L}_M} - 1)
\ln{\frac{r}{R_0}}},
\end{eqnarray} 
where $\sigma$ is the stellar velocity dispersion of the host galaxy,
and ${\cal L}$ is the quasar luminosity. $R_0$ is the launch radius of
the outflow, and $r$ the radius at which the velocity of the wind is
measured. ${\cal L}_M = \frac{4\ f_g\ c} {G} \sigma^4$ is a critical
luminosity that depends on the stellar velocity dispersion $\sigma$,
the speed of light, $c$, gravitational constant, $g$, and the gas
fraction, $f_g$. For ${\cal L}>{\cal L}_M$, radiation pressure may
launch a wind. These equations are appropriate for the limiting case
of an optically thick wind, in which case the interaction is most
efficient.

\citet{polletta08} estimated the stellar mass and luminosity of the
starburst and AGN from the dust and stellar emission of SW022513 with
an exquisite set of multi-waveband photometry. They find a stellar
mass of M$_{stellar}\sim 2-4\times 10^{11}$ M$_{\odot}$, and
bolometric luminosities of $6\times 10^{46} erg s^{-1}$ and $4.8\times
10^{46}$ erg s$^{-1}$ for the starburst and AGN, respectively.  For
pressure-supported galaxies with approximately isothermal mass
profile, this mass range corresponds to stellar velocity dispersions
of $\sigma= 300-350$ km s$^{-1}$. This can be found from setting M$=c
\sigma^2\ r_e / G$, where $\sigma$ is the stellar velocity dispersion,
$r_e$ the effective radius, $M$ the stellar mass, and G the
gravitational constant. For our calculations we adopted $r_e$ = 2-3
kpc, and $c=5$ \citep{bender92}. To give a lower limit on the gas
fraction, we use the molecular gas mass estimate of
\citet{polletta10}, $4\times 10^{10}$ M$_{\odot}$, which gives a gas
fraction of order f$_g= 0.1-0.2$ for a stellar mass of
M$_{stellar}=2-4\times 10^{11}$ M$_{\odot}$ \citep{polletta08}.

Following \citet{murray05}, to launch a wind in a galaxy with
M$_{stellar}=2 - 4 \times 10^{11}$ M$_{\odot}$ with
$\sigma$=300-350 km s$^{-1}$, and M$_{gas}=4\times 10^{10}$
M$_{\odot}$ would require ${\cal L}_M=2.5 - 3 \times 10^{47}$ erg
s$^{-1}$. This is the most optimistic case consistent with our
observational constraints, and would require a bolometric luminosity
that is about 2.5$-$3$\times$ greater than observed.  To accelerate a
wind to a terminal velocity of $\ge 1000$ km s$^{-1}$ (as suggested by
the broad blueshifted components in SW022513) would then require a
bolometric luminosity of at least $4.5-5\times 10^{47}$ erg s$^{-1}$
for a galaxy with $\sigma =$300$-$350 km s$^{-1}$, a factor 4-5 larger
than what is measured. Thus, the luminosity of SW022513 is at least
a factor 4 lower than required, if we consistently use the most
optimistic estimates implied by our observations. It can be up to
factors~10 lower for other plausible choices of parameters.

For these estimates we assume a launch radius of the wind, $R_0$, of a
few 100 pc \citep[the sizes of the circumnuclear molecular disks found
in low-redshift ULIRGs][ and the lowest value in the AGN feedback
models of \citealt{ciotti09,ciotti10}]{downes98}, and an outflow
radius, $r$, of 5~kpc, which roughly corresponds to the radius of the
narrow-line region in SW022513. A larger launch radius (perhaps more
plausible given the large gas masses) or a larger size of the outflow
region, enhances the required energy. This suggests that radiation
pressure may only drive the outflow if all parameters take values that
are strictly at their lower limits and if we underestimate the
luminosity by at least a factor~4. Otherwise, driving an outflow like
in SW022513 through radiation pressure as proposed by \citet{murray05}
would require an AGN and starburst with at least factors of a few
higher bolometric luminosity than in SW022513, although this is one of
the most luminous obscured high-z AGN in the SWIRE survey
\citep{polletta08}.

\subsubsection{Radio source}

Examples of outflows in the literature associated with radio-loud AGN
are numerous, and include not only powerful radio sources
\citep[e.g.,][]{morganti05, holt08, nesvadba08b, spoon09a, spoon09b,
nesvadba10a}, but also Seyfert galaxies with low radio power
\citep{capetti99, ulvestad99, reynolds09, barbosa09}. Weak radio
sources are common in Seyfert galaxies \citep{gallimore06} but not
always easily identified, and typically spatially asscociated with
non-circular gas motions. The same is found for radio-quiet quasars
\citep[e.g.,][]{leipski06a,leipski06b}. Note that ``radio-quiet'' does
not mean ``radio-silent'', but that the radio power is less than
10$\times$ greater than the optical luminosity of a quasar, 
irrespective of whether the optical continuum is dominated by the
direct or scattered AGN light, nebular continuum emission, or the
obscured or unobscured stellar continuum of the host galaxy.

Even relatively low-power radio sources can accelerate gas to high
velocities of $\ge$1000 km s$^{-1}$ \citep[e.g.][]{capetti99} and
produce equally large line widths \citep[e.g.,][]{capetti99,
reynolds09}. \citet{heckman81} found that radio power correlates with
the FWHM of the [OIII]$\lambda$5007 line. Measuring the
[OIII]$\lambda$5007 FWHM in SW022513 from the
integrated spectrum, we find that it falls well within the scatter of
the \citeauthor{heckman81} relationship
(Figure~\ref{fig:heckman}). The same is true for SW022550 and
SMMJ1237+6203 \citep[with the FWHM of][]{alexander10}

Weak radio sources where the internal pressure does not greatly exceed
the ambient pressure, may deposit most of their energy in the ambient
medium \citep{gallimore06} and may thus be more efficient in inflating
bubbles than more powerful radio sources,
where only a few percent of the kinetic jet energy is being deposited
in the ambient medium \citep[e.g.,][]{nesvadba08b}.
\citet{polletta08} estimated the radio power of SW022513 (and
SW022550) to be of order $10^{25}$ W Hz$^{-1}$. This roughly
corresponds to the FRI/FRII divide at low redshift, and appears as
{\it moderately} strong only if compared to the very powerful
radio-loud quasars and radio galaxies at high redshift. At more
moderate redshifts, radio sources with fairly similar power can
trigger significant outflows \citep[e.g.][]{morganti05,spoon09a} and
have a profound impact on the molecular gas in their hosts
\citep[e.g.,][]{papadopoulos10, nesvadba10a, ogle10, alatalo10}.

The interaction efficiency of a radio source with the ambient gas
depends critically on the gas density \citep[e.g.,][]{capetti99},
which in high-redshift galaxies is likely higher than at low redshift.
For example, \citet{deyoung93} suggest that trapping a radio source
with about $10^{25}$ W Hz$^{-1}$ in the ambient gas requires a
volume-averaged gas density of order 100 cm$^{-3}$. Assuming a
volume-averaged density may appear somewhat artificial, but has
previously shown to provide reasonable constraints
\citep{deyoung93,capetti99}, and is certaintly well matched to the
crudeness of our observational constraints.  For low-redshift
galaxies, plausible estimates of the volume-averaged ambient gas
density are of order 1-10 cm$^{-3}$ \citep{deyoung93}, however, a
value of 100 cm$^{-3}$ is very similar to the average density 50-380
cm$^{-3}$ in SW025513 and SW022550 implied by the molecular mass of
$4\times 10^{10}$ M$_{\odot}$ \citep{polletta10}, and assuming a
spherical gas distribution with radius 1-2 kpc and a filling
factor of unity. As a consequence, the higher gas densities in high
redshift galaxies could boost the effect of moderately strong radio
sources on their surrounding gas compared to nearby galaxies, where
such jets escape more easily.

In many similarly strong, low-redshift radio sources the total gas mass involved
in the outflow is much smaller than in SW022513
\citep[e.g.,][]{morganti05}. The H$\beta$ luminosity in the component with
FWHM$=$1000 km s$^{-1}$ implies an ionized gas mass of $1.2\times
10^{8-9}$ M$_{\odot}$, estimated with the method, assumptions, and
uncertainties presented in \S\ref{ssec:masses}. We therefore test
explicitly if the radio source in SW022513 may provide enough
mechanical energy to produce the observed emission-line kinematics of
H$\beta$ and CO. \citet{willott99} estimate the jet kinetic energy
from the
observed radio power ${\cal L}_{151,28}$ at 151 MHz, setting ${\cal
L}_{mech} = 3\times 10^{38}\ f_W^{3/2} {\cal L}_{151,28}^{6/7}$
W. ${\cal L}_{151,28}$ is the observed radio power at 151 MHz in units
of $10^{28}$ W Hz$^{-1}$ sr$^{-1}$, and $f_W$ is a fudge factor taking
into account the most salient astrophysical uncertainties. Typically
$f_W = 10$ \citep[see also][]{cattaneo09}. Estimating the 151 MHz
radio luminosity from the measured radio fluxes at 1.4 GHz and 610 MHz
\citep{polletta08}, we find a mechanical energy injection rate of
$2\times 10^{44}$ erg s$^{-1}$.

To estimate the total kinetic energy released by the radio source, we
need to estimate the AGN lifetime. \citet{blundell99} argue that the
most powerful radio sources at high redshift must be very young, about
$10^7$ yrs. We will use this estimate as a lower limit on the age of
the radio source. However, \citet{sajina07} find that 1/3$-$1/4 of
intense infrared-selected starbursts at z$\ge$2 have moderately
radio-loud AGN, with radio power and star-formation properties broadly
similar to our sources.  This suggests that moderately powerful
radio sources are not uncommon amongst intensely star forming galaxies
at high redshift. Assuming that all such galaxies are moderately
bright radio sources at some time of their evolution yields a lower
limit on the lifetime of radio sources in individual galaxies relative
to the lifetime of the starburst. For typical star-formation
timescales of few $10^8$ yrs, this would suggest that moderately
radio-loud AGN at z$\sim$2 have longer radio lifetimes than the most
extreme radio sources, up to of-order $10^8$ yrs, more resembling the
long activity periods of weak radio sources in nearby Seyfert galaxies
\citep{gallimore06} than the very short lifetimes (of-order $10^7$
yrs) of very powerful radio galaxies at high-z
\citep[e.g.,][]{blundell99}.

Even if the current energy injection rate of $2\times 10^{44}$ erg
s$^{-1}$ in SW022513 were only maintained for $1\times 10^7$ yrs, the
radio source would provide sufficient mechanical energy, $7\times
10^{58}$ erg, to accelerate large amounts of of molecular and ionized
gas as is observed, if all of the jet's mechanical energy is
transferred to the gas at a conversion efficiency of 100\%. For longer
timescales, the same could be achieved with lower conversion
efficiencies, of-order few 10\%, closer to those found in
very powerful radio galaxies at z$\ge$2 \citep{nesvadba08b}.

\subsection{Will the gas escape?}
\label{ssec:willthegasescape}
AGN-driven outflows were postulated by galaxy evolution models to
sweep up and unbind the remaining gas at the end of merger-triggered
major episodes of star formation at high redshift. Major gas-rich
mergers are the most plausible processes that may trigger intense star
formation and AGN activity in high-redshift galaxies, however, our
data do not reveal multiple continuum components for either source
(\S\ref{ssec:continuum}). This could imply that both components are at
distances less than the spatial resolution of our data ($\sim$4$-$5
kpc) or that their stellar components have low surface brightnesses,
perhaps because they are being tidally disrupted. Either interpretation
would be consistent with an advanced merger stage. The same is
suggested by the short gas consumption timescales of few $\times 10^7$
yrs, about 10\% of the age of the starburst \citep{polletta10}.
 
Finding outflows of molecular and ionized gas associated with an
obscured AGN is therefore certainly consistent with this broad
evolutionary picture. However, a critical question is whether the gas
will escape. In \S\ref{ssec:radiationpressure} we estimated that the
velocity dispersion corresponding to the stellar mass of SW022513
derived by \citet{polletta08} is likely about $\sigma=$300-350 km
s$^{-1}$, which would imply an escape velocity $\sqrt{2}\ \sigma \sim
500$ km s$^{-1}$.

Finding gas with velocities of up to 1000 km s$^{-1}$ certainly
suggests that at least a fraction of the outflow in gas may ultimately
become unbound, depending on how much energy is being dissipated by
the outflow before the gas has reached large radii. Recent
observations of molecular gas in nearby radio galaxies finding roughly
similar amounts of thermal and kinetic energy suggest that
dissipational processes in the molecular gas could be important for
the dynamics of molecular winds \citep{nesvadba10a}. The most direct
evidence for gas escaping from the galaxy would therefore be the
detection of outflowing gas that extends well beyond the radius of the
host galaxies, with large velocity offsets, and aligned with the radio
axis, as has been found in powerful radio galaxies at z$\ge$2
\citep{nesvadba06b, nesvadba07b, nesvadba08b, nesvadba10b}. At any
rate, our observations suggest that ``radio-quiet'', powerful AGN at
high redshift may accelerate significant fractions of their
interstellar medium to velocities near or above the escape velocity
even if their radio sources are rather unconspicuous.

\section{Summary}

We present an analysis of deep rest-frame optical integral-field
spectroscopy of two powerful obscured quasars at z$\ge$3.5, SW022513
and SW022550. These are the most luminous 24$\mu$m emitters in
the SWIRE survey at $\ge$2 and have previously been discussed by
\citet{polletta08,polletta10}. Our main results are as follows:

(1) The optical line emission in both galaxies is dominated by the
luminous narrow-line region ionized by the hard quasar
spectrum. Emission lines are very broad in SW022550, 
FWHM$=$2200 km s$^{-1}$ for [OIII]$\lambda$4959. Emission-line
profiles in SW022513 are complex. For example, [OIII]$\lambda$5007 is
dominated by a 'narrow' component with FWHM$=$1275 km s$^{-1}$ and has
a broad, blueshifted component of up to FWHM$=$5000 km s$^{-1}$. These
line widths reflect the kinematics of the narrow-line region, the
broad-line region is obscured.

(2) In SW022513 the line emission is spatially extended. We identify
the nucleus with the continuum peak and site of broadest
[OIII]$\lambda\lambda$4959,5007 line emission (FWHM$=$5000 km
s$^{-1}$). The peak in narrow [OIII]$\lambda\lambda$4959,5007 line
emission is offset by 0.75\arcsec\ to the South ($\sim$5 kpc). This
suggests that the ionized gas reaches the largest velocities near the
nucleus, and is surrounded by an extended narrow-line region. For
bright AGN like in SW022513 narrow-line region sizes of 10s of kpc are
possible. 

(3) For SW022513, and comparing with the CO(4--3) observations of
\citet{polletta10}, we find that the ionized gas mass amounts to
10\%-100\% of the molecular mass. CO(4--3) and H$\beta$ line profiles
are well matched suggesting both may originate from the same gas
clouds. Using the narrow H$\beta$ component near the nucleus to define
the systemic redshift (as commonly done in AGN host galaxies) we find
that about 40\% of the CO line emission is from gas that does not
participate in systemic motion, but is blueshifted to velocities of up
to $-$1000 km s$^{-1}$. Such large velocity offsets are not suggested
by merger models including starburst-driven winds and direct empirical
evidence, but could be a signature of AGN-driven winds as postulated,
e.g. by \citet{narayanan08}. In SW022550 the CO(4--3) line profile is
double-peaked and very different from that of [OIII], which may
suggest that the bulk of molecular gas is not affected by the AGN in a
similar way, although the FWHM of the [OIII] line of $>$2000 km
s$^{-1}$ suggests that the AGN interacts with the ISM nonetheless.

(4) SW022513 and SW022550 host AGN with bolometric luminosities of
$\sim 5 \times 10^{46}$ erg s$^{-1}$ and moderately strong radio
sources. Comparing with the expected characteristics of
radiation-pressure driven winds \citep[following][]{murray05} and
mechanical energy injection from the radio source
\citep[following][]{willott99} we find that it is difficult to produce
the observed velocities with radiation pressure, whereas observations
and basic energy considerations suggest the radio source as a possible
driver. In this case, moderately radio-loud ULIRGs like SW022513 and
SW022550 could be 'scaled-up' versions of nearby Seyfert galaxies and
'radio-quiet' (but not radio-silent) quasars with weak radio sources.

\section*{Acknowledgments}

We are very grateful to the staff at Paranal and at IRAM for having
carried out the observations on which our analysis is based. We also
thank the referee, Montserrat Villar-Martin, for comments that were
very helpful in improving the paper.

\bibliographystyle{mn2e}
\bibliography{hzrg}

\clearpage \onecolumn

\clearpage
\onecolumn

\clearpage 
\onecolumn

\clearpage
\onecolumn
\begin{figure}
\centering
\includegraphics[width=0.7\textwidth]{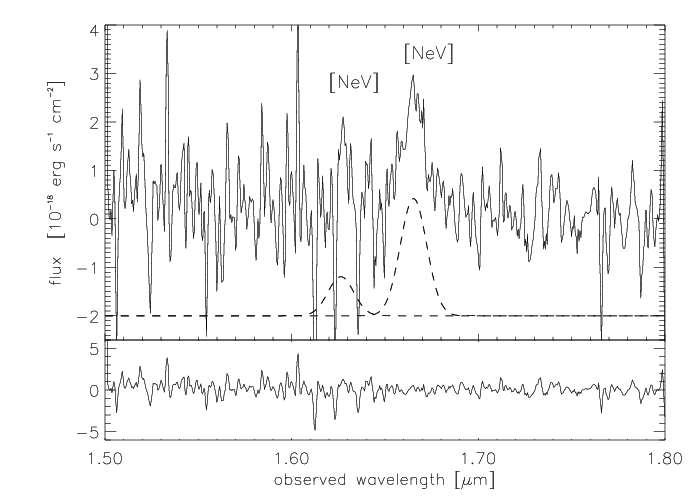}\\
\includegraphics[width=0.7\textwidth]{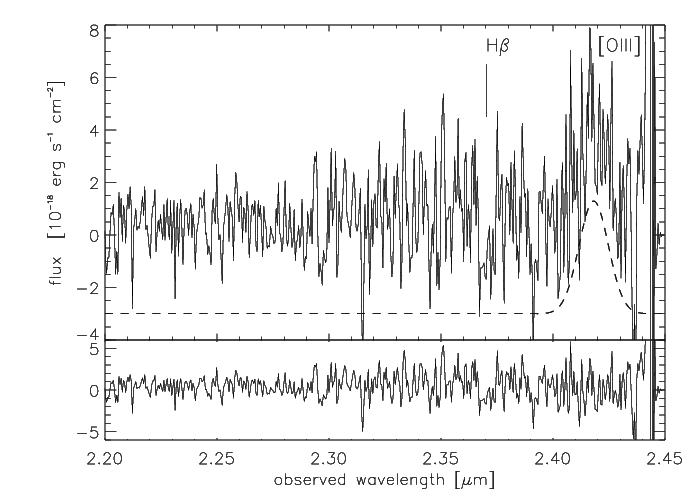}\\
\caption{Integrated spectrum of SW022550 showing
[NeV]$\lambda\lambda$3345,3426 (top panel) and [OIII]$\lambda$4959
(bottom panel).  [OIII]$\lambda$5007 falls outside of the K-band
atmospheric window, and H$\beta$ is not detected. Hatched lines mark
Gaussian line fits, and the bottom panels show the fit
residuals. [NeV]$\lambda$3345 is superimposed by several night
sky-line residuals, and we did not fit the line, but plot the line fit
with the redshift, and line width expected from those measured for the
[NeV]$\lambda$3426 line, and use a line ratio of R(3426,3345)$=$2.8
expected from the transition probabilities of each component of the doublet.}
\label{fig:SW022550_intspec}
\end{figure}

\clearpage 
\onecolumn
\begin{figure} \centering
\includegraphics[width=0.7\textwidth]{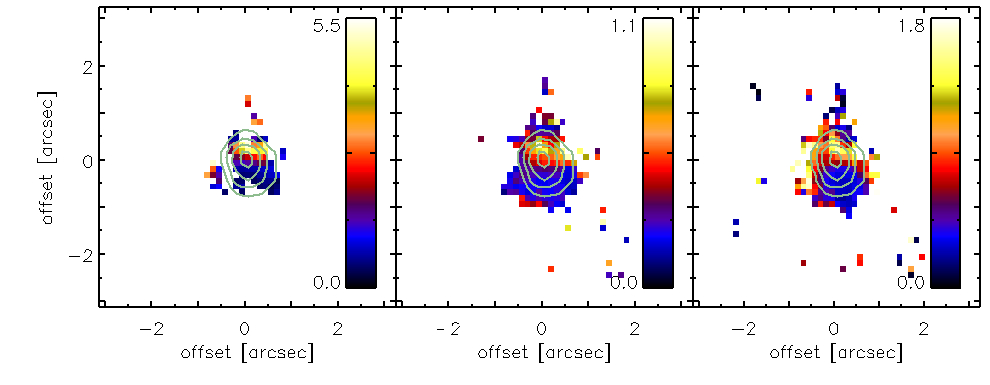}\\
\caption{Line ratios in SW022513 (left to right): Flux ratio of the broad
to narrow component of [OIII]$\lambda$5007. Ratio of H$\beta$ to
[OIII]$\lambda$5007 (narrow component). Ratio of
[OII]$\lambda\lambda$3726,3729 to [OIII]$\lambda$5007 (narrow
component). Both for H$\beta$ and [OII] line ratios relative to [OIII]
decrease with increasing distance from the AGN. Contours show the
continuum morphology in all panels, corresponding to the position of the AGN.}
\label{fig:SW022513_ratiomaps}
\end{figure}

\clearpage 
\onecolumn

\begin{figure} \centering
\includegraphics[width=0.7\textwidth]{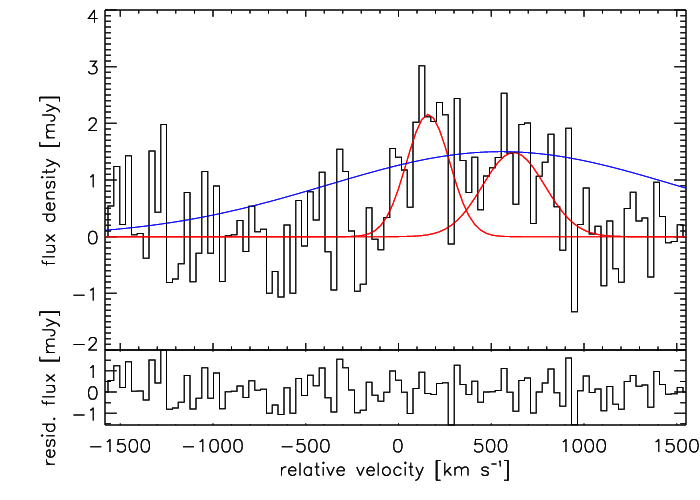}\\
\caption{Comparison of the line profile of CO(4--3) \citep[black
line][]{polletta10} and [OIII]$\lambda$4959 (blue line) in
SW022550. The lower panel shows the residual after fitting the CO
profile with two Gaussian distributions (which are shown in red 
in the upper panel). The [OIII]$\lambda$4959 profile was scaled
arbitrarily.}
\label{fig:SW022550_CO_Hbeta}
\end{figure}

\begin{figure} \centering
\includegraphics[width=0.7\textwidth]{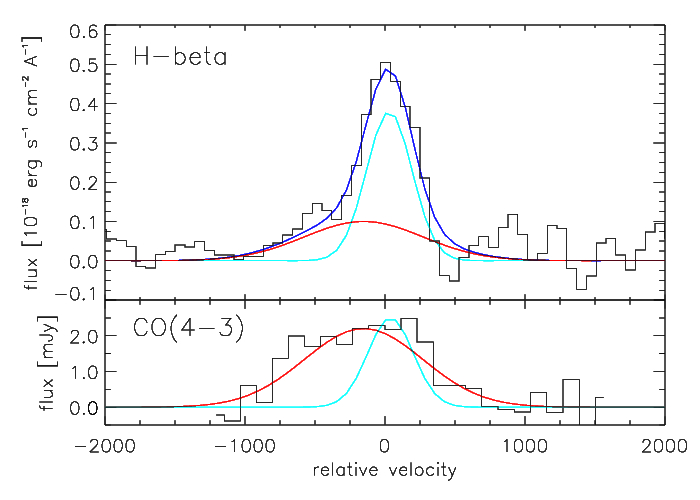}\\
\caption{Comparison of the line profile of H$\beta$ near the nucleus
(upper panel, see also the mid panel of
Figure~\ref{fig:SW022513_spectra}) and the CO(4--3) emission-line
spectrum of SW022513 (lower panel) presented by
\citet{polletta10}. The red, cyan, and blue Gaussian distributions
show the CO(4--3), narrow H$\beta$ component, and sum of both spectra,
respectively. The red and cyan distributions have the same wavelength
and width in both panels.}
\label{fig:SW022513_CO_Hbeta}
\end{figure}

\clearpage 
\onecolumn
\begin{figure} \centering
\includegraphics[width=0.7\textwidth]{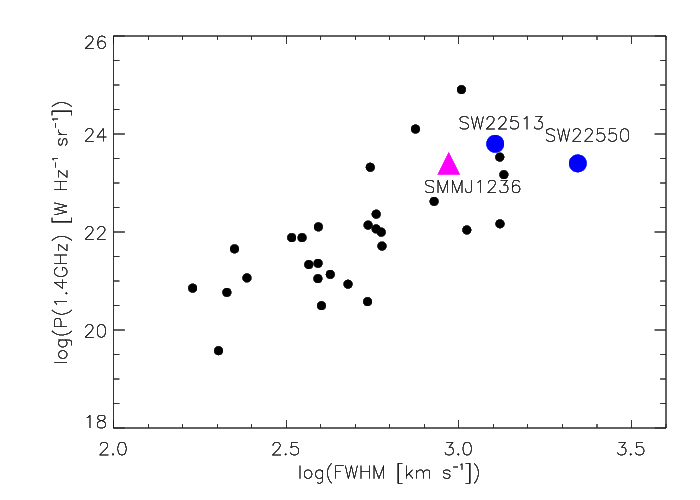}\\
\caption{Radio power measured at 1.4 GHz in the rest-frame as a
function of FWHM of the [OIII]$\lambda$5007 emission line. Small black
dots show nearby AGN spanning a large range in radio power, and are
taken from the original version of this plot shown in
\citet{heckman81}. Large blue dots show SW022550, and SW022513, the
red triangle shows SMMJ1237+6203 \citep{alexander10}.}
\label{fig:heckman}
\end{figure}

\clearpage
\onecolumn
\begin{table}
\centering
\begin{tabular}{rcccccc}
\hline
Line ID & $\lambda_0$ & z & $\lambda_{obs}$ & FWHM & FWHM & flux \\
(1)     & (2)         & (3)      & (4)            & (5)   & (6) & (7) \\
\hline
$[$OIII$]$  & 4959        & 3.876$\pm$0.001 & 24182$\pm$7 & 179.5 & 2212$\pm$178 & 8.2$\pm$0.76 \\
$[$NeV$]$   & 3456        & 3.861$\pm$0.002 & 16653$\pm$6 & 166.3 & 2985$\pm$267 & 4.28$\pm$45 \\
\hline

\end{tabular}
\caption{Emission-line properties in the integrated spectrum of
SW022550 Column (1) -- Line ID. Column (2) -- Rest-frame wavelength in
\AA. Column (3) -- Redshift. Column (4) -- Observed wavelength in
\AA. Column (5) -- Observed line width (Full Width at Half Maximum) in
\AA, not corrected for instrumental resolution. Column (6) -- FWHM
corrrected for instrumental resolution in km s$^{-1}$. Integrated
emission-line flux in 10$^{-16}$ erg s$^{-1}$ cm$^{-2}$.}
\label{tab:spectraSW022550}
\end{table}

\begin{table}
\centering
\begin{tabular}{rcccccc}
\hline
Line ID & $\lambda_0$ & z & $\lambda_{obs}$ & FWHM & FWHM & flux \\
(1)     & (2)         & (3)      & (4)            & (5)   & (6) & (7) \\
\hline
$[$OIII$]$ & 5007 & 3.42312$\pm$0.0009 & 221466$\pm$4 & 95.9$\pm$6  & 1275$\pm$77    & 40.9$\pm$2.5 \\
$[$OIII$]$ & 4959 & 3.44239$\pm$0.0005 & 219380$\pm$5 & 93.9$\pm$6  & 1259$\pm$82    & 14.6$\pm$1.0 \\ 
H$\beta$  & 4861 & 3.42438$\pm$0.0009  & 215069$\pm$5 & 69.1$\pm$5  & 929.6$\pm$70   & 6.8$\pm$0.6 \\
$[$OII$]$ & 3727 & 3.42480$\pm$0.0009  & 164912$\pm$3 & 105.1$\pm$7 & 1895.1$\pm$120 & 19.4$\pm$0.1 \\ 
\hline
\end{tabular}
\caption{Emission-line properties in the integrated spectrum of
SW022513 Column (1) -- Line ID. Column (2) -- Rest-frame wavelength in
\AA. Column (3) -- Redshift. Column (4) -- Observed wavelength in
\AA. Column (5) -- Observed line width (Full Width at Half Maximum) in
\AA, not corrected for instrumental resolution. $[$OII$]$ was fitted
with a single component at a rest-frame wavelength of 3727\AA. Column
(6) -- FWHM corrrected for instrumental resolution in km
s$^{-1}$. Integrated emission-line flux in 10$^{-16}$ erg s$^{-1}$
cm$^{-2}$.}
\label{tab:spectraSW022513}
\end{table}

\begin{table}
\centering
\begin{tabular}{rcccccc}
\hline
Line ID & $\lambda_0$ & z & $\lambda_{obs}$ & FWHM & FWHM & flux \\
(1)     & (2)         & (3)      & (4)            & (5)   & (6) & (7) \\
\hline
$[$OIII$]$ & 5007& 3.4227$\pm$0.0009 & 22144$\pm$4  & 46$\pm$3  & 579$\pm$40    & 1.22$\pm$0.09 \\
$[$OIII$]$ & 4959& 3.4227$\pm$0.0010 & 21932$\pm$5  & 45$\pm$5  & 579$\pm$70    & 0.40$\pm$0.06 \\
$[$OIII$]$ & 5007& 3.4032$\pm$0.0018 & 22047$\pm$9  & 374$\pm$32& 5087$\pm$447  & 2.58$\pm$0.25 \\
$[$OIII$]$ & 4959& 3.4031$\pm$0.0058 & 21835$\pm$28 & 370$\pm$95& 5087$\pm$1309 & 0.85$\pm$0.25 \\
H$\beta$   & 4861& 3.4247$\pm$0.0010 & 21504$\pm$5  & 26$\pm$5  & 369$\pm$40    & 0.32$\pm$0.05 \\
H$\beta$   & 4861& 3.422 $\pm$0.0010 & 21491$\pm$5  & 72$\pm$15 & 1000$\pm$73   & 0.30$\pm$0.07 \\
$[$OII$]$  & 3727& 3.4240$\pm$0.0010 & 16488$\pm$3  & 58$\pm$4  & 1037$\pm$73   & 1.19$\pm$0.09 \\
$[$NeIII$]$& 3869& 3.4248$\pm$0.0010 & 17117$\pm$4  & 41$\pm$4  & 682$\pm$82    & 0.37$\pm$0.06 \\
\hline
\end{tabular}
\caption{Emission-line properties near the continuum peak of SW022513
extracted from a 0.4\arcsec$\times$0.4\arcsec\ aperture, including the
broad and narrow components of $[$OIII$]$. Column (1) -- Line
ID. Column (2) -- Rest-frame wavelength in \AA. Column (3) --
Redshift. Column (4) -- Observed wavelength in \AA. Column (5) --
Observed line width (Full Width at Half Maximum) in \AA, not corrected
for instrumental resolution. Column (6) -- FWHM corrrected for
instrumental resolution in km s$^{-1}$. Integrated emission-line flux
in 10$^{-16}$ erg s$^{-1}$ cm$^{-2}$. We adopt the redshift of the
narrow H$\beta$ component as systemic redshift for SW022513 (see
text).}
\label{tab:spectraSW022513_contpeak}
\end{table}

\begin{table}
\centering
\begin{tabular}{rcccccc}
\hline
Line ID & $\lambda_0$ & z & $\lambda_{obs}$ & FWHM & FWHM & flux \\
(1)     & (2)         & (3)      & (4)            & (5)   & (6) & (7) \\
\hline
$[$OIII$]$  & 5007 & 3.4231$\pm$0.0009& 22146$\pm$4  & 37$\pm$2  &   455$\pm$29 & 3.34$\pm$0.22 \\
$[$OIII$]$  & 4959 & 3.4231$\pm$0.0009& 21934$\pm$5  & 37$\pm$3  &   455$\pm$38 & 1.10$\pm$0.1  \\
$[$OIII$]$  & 5007 & 3.4143$\pm$0.0016& 22102$\pm$8  & 159$\pm$19& 2152$\pm$257 & 1.34$\pm$0.2  \\
$[$OIII$]$  & 4959 & 3.4143$\pm$0.0059& 21890$\pm$29 & 158$\pm$66& 2152$\pm$907 & 0.44$\pm$0.23.\\
H$\beta$    & 4861 & 3.4238$\pm$0.0010& 21505$\pm$5  & 43$\pm$5  &   553$\pm$70 & 0.64$\pm$0.1  \\
$[$OII$]$   & 3727 & 3.4245$\pm$0.0009& 16490$\pm$4  & 42$\pm$3  &   734$\pm$65 &  0.93$\pm$0.1 \\
$[$NeIII$]$ & 3869 & 3.4254$\pm$0.0012&  17121$\pm$5 & 34$\pm$8  &   559$\pm$136& 0.27$\pm$0.08 \\
\hline
\end{tabular}
\caption{Emission-line properties near the peak of the
[OIII]$\lambda$5007 line emission in SW022513 extracted from a
0.4\arcsec$\times$0.4\arcsec\ aperture, including the broad and narrow
components of $[$OIII$]$. Column (1) -- Line ID. Column (2) --
Rest-frame wavelength in \AA. Column (3) -- Redshift. Column (4) --
Observed wavelength in \AA. Column (5) -- Observed line width (Full
Width at Half Maximum) in \AA, not corrected for instrumental
resolution. Column (6) -- FWHM corrrected for instrumental resolution
in km s$^{-1}$. Integrated emission-line flux in 10$^{-16}$ erg
s$^{-1}$ cm$^{-2}$.}
\label{tab:spectraSW022513_OIIIpeak}
\end{table}

\begin{table}
\centering
\begin{tabular}{rcccccc}
\hline
Line ID & $\lambda_0$ & z & $\lambda_{obs}$ & FWHM & FWHM & flux \\
(1)     & (2)         & (3)      & (4)            & (5)   & (6) & (7) \\
\hline
$[$OIII$]$ & 5007 & 3.4253$\pm$0.0009& 22157$\pm$5& 47$\pm$3& 604$\pm$42 & 1.43$\pm$0.11\\
H$\beta$   & 4861 & 3.4254$\pm$0.0009& 21513$\pm$4& 20$\pm$2& 165$\pm$23 & 0.29$\pm$0.05\\
$[$OII$]$  & 3727 & 3.4262$\pm$0.0010& 16496$\pm$4& 35$\pm$3& 606$\pm$65 & 1.12$\pm$0.15\\
$[$NeIII$]$& 3869 & 3.4273$\pm$0.0010& 17128$\pm$4& 27$\pm$5& 409$\pm$76 & 0.53$\pm$0.13\\
\hline
\hline
\end{tabular}
\caption{Emission-line properties in the elongated northern part of
SW022513 extracted from a 0.7\arcsec$\times$0.7\arcsec\
aperture. Column (1) -- Line ID. Column (2) -- Rest-frame wavelength
in \AA. Column (3) -- Redshift. Column (4) -- Observed wavelength in
\AA. Column (5) -- Observed line width (Full Width at Half Maximum) in
\AA, not corrected for instrumental resolution. Column (6) -- FWHM
corrrected for instrumental resolution in km s$^{-1}$. Integrated
emission-line flux in 10$^{-16}$ erg s$^{-1}$ cm$^{-2}$.}
\label{tab:spectraSW022513_north}
\end{table}

\end{document}